\documentclass{aims} 
\usepackage{amsmath}
\usepackage{paralist}
\usepackage[misc]{ifsym}
  \usepackage{bm}
\usepackage[pagewise]{lineno}
\usepackage{epsfig} 
\usepackage{epstopdf} 
\usepackage[colorlinks=true]{hyperref}
\hypersetup{urlcolor=blue, citecolor=red}
\allowdisplaybreaks

\def\currentvolume{X}
 \def\currentissue{X}
  \def\currentyear{200X}
   \def\currentmonth{XX}
    \def\ppages{X-XX}
     \def\DOI{10.3934/xx.xxxxxxx}

\newtheorem{theorem}{Theorem}[section]

\theoremstyle{definition}

\newtheorem{remark}[theorem]{Remark}

\title[Summational invariant and local Maxwellian]
{Note on the summational invariant and corresponding local Maxwellian
      for the Enskog equation} 

\author[Shigeru TAKATA and Aoto TAKAHASHI]{}

\subjclass{Primary: 82C40, 82D05; Secondary: 76P05.}
\keywords{summational invariant, local Maxwellian, Enskog equation, kinetic theory, dense gas.}

\thanks{$^*$Corresponding author: Shigeru TAKATA}

\begin{document}
\maketitle

\centerline{\scshape
Shigeru TAKATA$^{{\href{mailto:takata.shigeru.4a@kyoto-u.ac.jp}{\textrm{\Letter}}}*1}$
and Aoto TAKAHASHI$^{{\href{mailto:takahashi.aoto.63c@st.kyoto-u.ac.jp}{\textrm{\Letter}}}1}$}

\medskip

{\footnotesize
 \centerline{$^1$Department of Aeronautics and Astronautics, Kyoto University, Kyoto 615-8540, Japan}
} 

%

\bigskip

 \centerline{(Communicated by Handling Editor)}


\begin{abstract}
The summational invariant and the corresponding local Maxwellian that are compatible with the Enskog equation are discussed,
with special interest in the presence of a boundary.
The local Maxwellian corresponding to the summational invariant is restrictive
compared to the case of the Boltzmann equation 
in the sense that a radial flow and time-dependent temperature are forbidden.
However, a rigid body rotation with a constant angular velocity is admitted 
as in the case of the Boltzmann equation.
The influence of the presence of a boundary is also discussed in simple situations.
\end{abstract}


\section{Introduction}
It is widely accepted that the Boltzmann equation describes the ideal gas behavior well 
for the entire range of the Knudsen numbers, the ratio of the mean free path of
gas molecules to a characteristic length of the system. The Boltzmann equation
is the most fundamental equation in the kinetic theory,
which today has a wide range of applications, such as {chemically reacting gases, dense gases, granular gases, traffic flows, electric transports in semiconductors, collective motions of chemotactic bacteria}.
The extension of the Boltzmann equation to a dense gas is one of the most classical ones, dating back to the work by Enskog \cite{E72}.
He proposed a kinetic equation, now called the (original) Enskog equation,
that takes into account the different center positions and correlations of molecules 
in the collision integral for a hard-sphere gas.
Despite its satisfactory results on the transport properties of a dense gas \cite{CC95,HCB64}
followed by successful applications to fundamental flows (e.g., \cite{F97b,F99,WLRZ16,HTT22}),
the original Enskog equation encountered the difficulty
of proving the H theorem, which had been the cornerstone of the kinetic theory since Boltzmann.
This difficulty stimulated further research \cite{R78,VE73,DVK21} on the Enskog equation
and gave rise to its variants.
To date, the H theorem has been proved in two cases:
(i) {correlation} of molecules is neglected, i.e., the so-called Boltzmann--Enskog equation \cite{BLPT91,DVK21,HN06}; and (ii) {correlation} of molecules is more complicated than {in} the original Enskog equation, i.e., the so-called modified Enskog equation \cite{VE73,R78}. 

For a long time, theoretical studies on the Enskog equation were mostly 
concerned with a gas in a periodic domain or with an infinite expanse of gas.
However, as pointed out in \cite{MGB18}, the finite-size effect of molecules in the collision integral makes the dynamics of a system with physical boundary more involved
than that of a system without boundary, requiring additional considerations even in simple situations (see also a numerical example in \cite{F97}). In the present paper, motivated by \cite{MGB18,F97}, we revisit the summational invariant and the corresponding local Maxwellian (or Maxwell distribution) that are compatible with the Enskog equation in a system with and without physical boundary. In a system with boundary, even the summational invariant needs a special care near the boundary, since a part of the contact directions of two colliding molecules is forbidden. Nevertheless, we are not aware of treating this problem in the literature, except for a well-prepared analysis between parallel plates 
 for the Boltzmann--Enskog equation by Brey \textit{et al.} \cite{BGM17}. 

{After} a brief preparation in Sec.~\ref{sec:EE}, we discuss the summational invariant in Sec.~\ref{sec:SI} by adapting Boltzmann’s original arguments \cite{B95} to the case 
with a restriction on the direction of contact.
Then, in Sec.~\ref{sec:LM}, we consider the local Maxwellian along the lines of Grad's argument for the Boltzmann equation \cite{G49,S07}.
We will show by an elementary argument that
the local Maxwellian representing a rigid body rotation is admissible  
for the Enskog equation, although it was excluded in the seminal paper of Resibois \cite{R78}.
The rigid body rotation mode of the Maxwellian is numerically demonstrated in Sec.~\ref{sec:ND}.
The paper is concluded in Sec.~\ref{sec:conclusion}.

\section{The Enskog equation\label{sec:EE}}
Let $D$ be a fixed spatial domain that the centers of gas molecules can occupy, where $D$ may be unbounded or bounded by a physical boundary. 
Let $t$, $\bm{X}$ and $\bm{Y}$, and $\bm{\xi}$
be { time}, spatial positions, and { molecular velocity}, respectively.
Then, denoting the one-particle distribution function of gas molecules
by $f(t,\bm{X},\bm{\xi}$) and the correlation function 
by $g(t,\bm{X},\bm{Y})$, the Enskog equation is written
as \begin{subequations}\label{MEE}
\begin{align}
 & \frac{\partial f}{\partial t}+\xi_{i}\frac{\partial f}{\partial X_{i}}=J(f)\equiv J^{G}(f)-J^{L}(f),\quad \mathrm{for\ }\bm{X}\in D,\displaybreak[0]\label{eq:2.1}\\
 & J^{G}(f)\equiv\frac{\sigma^{2}}{m}\int g(\bm{X}_{\sigma\bm{\alpha}}^{+},\bm{X})f_{*}^{\prime}(\bm{X}_{\sigma\bm{\alpha}}^{+})f^{\prime}(\bm{X})V_{\alpha}\theta(V_{\alpha})d\Omega(\bm{\alpha})d\bm{\xi}_{*},\displaybreak[0]\label{eq:2.2}\\
 & J^{L}(f)\equiv\frac{\sigma^{2}}{m}\int g(\bm{X}_{\sigma\bm{\alpha}}^{-},\bm{X})f_{*}(\bm{X}_{\sigma\bm{\alpha}}^{-})f(\bm{X})V_{\alpha}\theta(V_{\alpha})d\Omega(\bm{\alpha})d\bm{\xi}_{*},\label{eq:2.3}
\end{align}
\end{subequations}
\noindent where $\sigma$ and $m$ are {the} diameter and {the} mass of a molecule,
$\bm{X}_{\bm{x}}^{\pm}=\bm{X}\pm\bm{x}$,
$\bm{\alpha}$ is a unit vector, 
\begin{equation}
\theta(x)=\begin{cases}
1, & x\ge0\\
0, & x<0
\end{cases},
\end{equation}
\noindent $d\Omega(\bm{\alpha})$ is a
solid angle element in the direction of $\bm{\alpha}$,
and the following notation convention is used:
\begin{align}
 & \begin{cases}
f(\bm{X})=f(\bm{X},\bm{\xi}),\ f^{\prime}(\bm{X})=f(\bm{X},\bm{\xi}^{\prime}),\\
f_{*}(\bm{X}_{\sigma\bm{\alpha}}^{-})=f(\bm{X}_{\sigma\bm{\alpha}}^{-},\bm{\xi}_{*}),\ f_{*}^{\prime}(\bm{X}_{\sigma\bm{\alpha}}^{+})=f(\bm{X}_{\sigma\bm{\alpha}}^{+},\bm{\xi}_{*}^{\prime}),
\end{cases}\displaybreak[0]\\
 & \bm{\xi}^{\prime}=\bm{\xi}+V_{\alpha}\bm{\alpha},\quad\bm{\xi}_{*}^{\prime}=\bm{\xi}_{*}-V_{\alpha}\bm{\alpha},\quad V_{\alpha}=\bm{V}\cdot\bm{\alpha},\quad\bm{V}=\bm{\xi_{*}}-\bm{\xi}.\label{eq:2.5}
\end{align}
\noindent 
{The range of integrations in \eqref{eq:2.2} and \eqref{eq:2.3} is over the entire range of $\bm{\xi}_*$ and all directions of $\bm{\alpha}$.}
Here and in what follows, the argument $t$ is suppressed, unless confusion is anticipated.
Our correlation function $g$ is adjusted to the domain $D$ in such a way that the usual correlation function $g_{2}(t,\bm{X},\bm{Y})$ is modified as
\begin{subequations}\label{g-g2}\begin{align}
 & g(t,\bm{X},\bm{Y})=g_{2}(t,\bm{X},\bm{Y})\chi_{D}(\bm{X})\chi_{D}(\bm{Y}),\displaybreak[0]\label{eq:g_g2}\\
 & \chi_{D}(\bm{X})=\begin{cases}
1, & \bm{X}\in D\\
0, & \mbox{otherwise}
\end{cases},
\end{align}\end{subequations}
\noindent where $\chi_D$ plays the same role as the Heaviside function $\theta$,
when $D$ is bounded.
Among the variants of the Enskog equation,
the H theorem is proved for the Boltzmann--Enskog equation and for the modified Enskog equation, but not for the original Enskog equation.
Their difference is in the form of $g_2$.
The Boltzmann--Enskog equation is the simplest and $g_2=1$.
The original Enskog equation is more complicated, but $g_2$ is to some extent given freely as a function of a gas density, see, e.g., \cite{E72,HCB64,F97}.
The modified Enskog equation \cite{VE73,DVK21} is the most involved
and the expression of $g_2$ is not straightforward, 
see, e.g., \cite{R78,MGB18,T23}.
Fortunately, however, these differences are not relevant in the present paper.
Here, we just state that
$g_{2}$ has a symmetric property $g_{2}(t,\bm{X},\bm{Y})=g_{2}(t,\bm{Y},\bm{X})$
and a functional of a gas density 
\begin{equation}
\rho=\int fd\bm{\xi}.\label{eq:density}
\end{equation}
\noindent Thus, \eqref{MEE} is a closed equation for $f$
and will be referred to simply as the Enskog equation,
unless the above distinction is necessary.
By \eqref{g-g2}, $g$ has
the same symmetric property as $g_{2}$:
\begin{equation}
g(t,\bm{X},\bm{Y})=g(t,\bm{Y},\bm{X}).
\end{equation}


The summational invariant in the context of the Enskog equation
arises in the course of analysis of the H theorem \cite{R78,HN06,MGB18}
{ (see Sec.~4 of \cite{T23} for details)
and is defined by the following relation that holds in a stationary state:
\begin{equation}
 \ln f_{*}^{\prime}(\bm{X}_{\sigma\bm{\alpha}}^{-})+\ln f^{\prime}(\bm{X})
=\ln f_{*}(\bm{X}_{\sigma\bm{\alpha}}^{-})+\ln f(\bm{X}),
\label{eq:S1}
\end{equation}
\noindent for the entire range of $\bm{\xi}$ and $\bm{\xi}_*$ and for $\bm{X}, \bm{X}_{\sigma\bm{\alpha}}^{-}\in D$.}
The quantity $\ln f$ above is what we call the summational invariant.
The difference from the case of the Boltzmann equation is that a finite-size effect of molecules appears in \eqref{eq:S1}. 

\section{Summational invariant\label{sec:SI}}
Because of the restriction $\bm{X}^-_{\sigma\bm{\alpha}}\in D$,
\eqref{eq:S1} does not have to hold for a part of directions of $\bm{\alpha}$,
if $\bm{X}$ is near the boundary $\partial D$.
We will seek a general form of $\ln f$ that satisfies \eqref{eq:S1} for the entire space of $(\bm{\xi},\bm{\xi}_*)$ with $\bm{\alpha}$ being fixed.
This is a main difference from the standard proofs for the Boltzmann equation, e.g., \cite{K38,G49}.
Once $\bm{\alpha}$ is fixed, the sub-domain of $D$ that $\bm{X}$ can occupy is fixed.
Then, we follow, to some extent, Boltzmann's original idea
for his own equation \cite{B95} 
that makes use of the Lagrange multiplier method and treats all velocities
$\bm{\xi}$, $\bm{\xi}_*$, $\bm{\xi}^\prime$, and $\bm{\xi}_*^\prime$ as independent  variables.

Consider the variation of \eqref{eq:S1} with respect to $\bm{X}$, $\bm{\xi}$, $\bm{\xi}_*$, $\bm{\xi}^\prime$, $\bm{\xi}_*^\prime$, as if they were all independent.
Actually, however, among $3+3\times 4=15$ variables, there are only $3+3\times2=9$ independent variables. In other words, there are six constraints arising from the momentum, the energy, and the angular momentum conservation:%
\footnote{There are actually only two independent equations in \eqref{eq:angle}. In accordance with this redundancy, three undetermined constants denoted by $\bm{\gamma}$ appear soon later.}
\begin{subequations}\label{eq:constraints}\begin{align}
  \bm{\xi}+\bm{\xi}_*
&=\bm{\xi}^\prime+\bm{\xi}_*^\prime,\label{eq:momentum}\displaybreak[0]\\
   \bm{\xi}^2+\bm{\xi}_*^2
&={\bm{\xi}^\prime}^2+{\bm{\xi}_*^\prime}^2,\label{eq:energy}\displaybreak[0]\\
  \bm{X}\times\bm{\xi}+\bm{X}_{\sigma\bm{\alpha}}^-\times\bm{\xi}_*
&=\bm{X}\times\bm{\xi}^\prime+\bm{X}_{\sigma\bm{\alpha}}^-\times\bm{\xi}_*^\prime.\label{eq:angle}
\end{align}\end{subequations}
\noindent Taking the variations of \eqref{eq:S1} and \eqref{eq:constraints}
and using the Lagrange multiplier method, the following identities are obtained:
\begin{subequations}\label{eq:variation}
\begin{align}
&  \frac{\partial\ln f(\bm{X})}{\partial \bm{\xi}}-\bm{\lambda}-\bm{\gamma}\times\bm{X}-2\mu\bm{\xi}=0,\displaybreak[0]\\ 
&  \frac{\partial\ln f_*(\bm{X}_{\sigma\bm{\alpha}}^-)}{\partial \bm{\xi}_{*}}-\bm{\lambda}-\bm{\gamma}\times\bm{X}_{\sigma\bm{\alpha}}^--2\mu\bm{\xi}_{*} =0,\label{eq:var-vel}\displaybreak[0]\\
&  \frac{\partial\ln f^\prime(\bm{X})}{\partial \bm{\xi}^\prime}-\bm{\lambda}-\bm{\gamma}\times\bm{X}-2\mu\bm{\xi}^\prime=0,\displaybreak[0]\\ 
&  \frac{\partial\ln f^\prime_*(\bm{X}_{\sigma\bm{\alpha}}^-)}{\partial \bm{\xi}_{*}^\prime}-\bm{\lambda}-\bm{\gamma}\times\bm{X}_{\sigma\bm{\alpha}}^--2\mu\bm{\xi}_{*}^\prime =0,\label{eq:var-vel2}
\end{align}\end{subequations}
\noindent and
\begin{align}
 \frac{\partial}{\partial \bm{X}}
& \{\ln f(\bm{X})+\ln f_*(\bm{X}_{\sigma\bm{\alpha}}^-)
   -\ln f^\prime(\bm{X})
   -\ln f_*^\prime(\bm{X}_{\sigma\bm{\alpha}}^-)\}\notag\\
&+\bm{\gamma}\times(\bm{\xi}+\bm{\xi}_{*}-\bm{\xi}^\prime-\bm{\xi}_{*}^\prime) =0,
\label{eq:var-pos}
\end{align}
\noindent where $\bm{\lambda}$, $\bm{\gamma}$, and $\mu$ are undetermined multipliers.
Integrating \eqref{eq:variation} with respect to the molecular velocity yields
\begin{equation}
\ln f(\bm{X})=(\bm{\lambda}-\bm{X}\times\bm{\gamma})\cdot\bm{\xi}+\mu\bm{\xi}^2+\beta(\bm{X}),
\label{eq:sum}
\end{equation}
\noindent where $\beta(\bm{X})$ is a constant of integration, and substituting \eqref{eq:sum} into \eqref{eq:var-pos} shows that $\beta(\bm{X})$ is arbitrary.
Since the dependence on time $t$ has been suppressed in the above discussion,
$\bm{\lambda}$, $\bm{\gamma}$, $\mu$, and $\beta(\bm{X})$ may depend on $t$ in general.
This is consistent with the form given in \cite{MGB18} and  
is more restrictive than the case of the Boltzmann equation. The restriction originates from the difference of centers of two colliding molecules. See also Appendix~\ref{sec:app}.

\begin{remark}
We have implicitly assumed that $\ln f$ is differentiable and there is a subdomain of $D$
where any direction of $\bm{\alpha}$ can be taken.
In the former sense, our approach is similar,
though not identical, to that in \cite{BGM17}.
For the Boltzmann equation, a general form of the summational invariant is obtained under a weaker assumption, see, e.g., \cite{TM80,CIP94,P98}.
To our knowledge, the applicability of the methods in these references has not yet been examined.
\end{remark}

\section{Local Maxwellian\label{sec:LM}}
Because of the form \eqref{eq:sum}, 
the summational invariant requires that
the corresponding velocity distribution function $f_e$ 
is the local Maxwellian in the form
\begin{subequations}\label{eq:LocalMax}\begin{equation}\label{eq:Max}
f_e=\frac{\rho(t,\bm{X})}{(2\pi RT(t))^{3/2}}
    \exp(-\frac{(\bm{\xi}-\bm{v}(t,\bm{X}))^2}{2RT(t)}),
\end{equation}
\noindent where
\begin{equation}
\bm{v}(t,\bm{X})=\bm{u}(t)+\bm{X}\times\bm{\omega}(t),
\label{eq:velocity}
\end{equation}\end{subequations}
and the following correspondence among quantities occurring in \eqref{eq:sum} and \eqref{eq:LocalMax} should be reminded:
\begin{equation}
\beta=\ln\frac{\rho}{(2\pi RT)^{3/2}}-\frac{\bm{v}^2}{2RT},
\quad
\mu=-\frac{1}{2RT},
\quad \bm{\lambda}=\frac{\bm{u}}{RT},
\quad \bm{\gamma}=-\frac{\bm{\omega}}{RT}.\label{eq:Lag}
\end{equation}
\noindent {
Note that $\bm{u}$, $\bm{\omega}$, and $T$ are also independent of $\bm{X}$
because $\bm{\lambda}$, $\bm{\gamma}$, and $\mu$ are independent of $\bm{X}$.}

Let us now substitute \eqref{eq:Max} into the Enskog equation \eqref{MEE}
\begin{equation}
 \frac{\partial f_e}{\partial t}
+\xi_i\frac{\partial f_e}{\partial X_i}
=J(f_e),\label{ES}
\end{equation}
\noindent and examine \eqref{ES} along the lines of Grad's discussions
\cite{G49,S07} on the Boltzmann equation.
The main difference from the Boltzmann equation is that
$J(f_e)$ does not vanish in general. 
Indeed, it is reduced only to
\begin{equation}
 J(f_e)
=-\frac{\sigma^2}{m}f_e(\bm{X})(\bm{\xi}-\bm{v}(\bm{X}))\cdot
\int \bm{\alpha} g(\bm{X}_{\sigma\bm{\alpha}}^+,\bm{X})
\rho(\bm{X}_{\sigma\bm{\alpha}}^+) d\Omega(\bm{\alpha}),\label{eq:RHS}
\end{equation}
\noindent which is shown as follows. First note that
\begin{equation}
 f_{e*}^{\prime}(\bm{X}_{\sigma\bm{\alpha}}^{+})f_e^{\prime}(\bm{X})
=f_{e*}(\bm{X}_{\sigma\bm{\alpha}}^{+})f_e(\bm{X}),
\end{equation}
\noindent since 
\begin{align}
 & (\bm{\xi}_*^\prime-\bm{v}(\bm{X}_{\sigma\bm{\alpha}}^+))^2
  +(\bm{\xi}^\prime-\bm{v}(\bm{X}))^2 \notag\displaybreak[0]\\
=& (\bm{\xi}_*^\prime-\bm{v}(\bm{X})-\Delta\bm{v})^2
  +(\bm{\xi}^\prime-\bm{v}(\bm{X}))^2 \notag\displaybreak[0]\\
=& (\bm{\xi}_*^\prime-\bm{v}(\bm{X}))^2-2(\bm{\xi}_*^\prime-\bm{v}(\bm{X}))\cdot\Delta\bm{v}+(\Delta\bm{v})^2 +(\bm{\xi}^\prime-\bm{v}(\bm{X}))^2 \notag\displaybreak[0]\\
=& (\bm{\xi}_*-\bm{v}(\bm{X}))^2-2(\bm{\xi}_*-\bm{v}(\bm{X}))\cdot\Delta\bm{v}+(\Delta\bm{v})^2 +(\bm{\xi}-\bm{v}(\bm{X}))^2 \notag\displaybreak[0]\\
=& (\bm{\xi}_*-\bm{v}(\bm{X}_{\sigma\bm{\alpha}}^+))^2 +(\bm{\xi}-\bm{v}(\bm{X}))^2.
\label{eq:20}
\end{align}
\noindent Here the identities in \eqref{eq:20} come from the facts that
(i) $\Delta\bm{v}\equiv\bm{v}(\bm{X}_{\sigma\bm{\alpha}}^+)-\bm{v}(\bm{X})=\sigma\bm{\alpha}\times\bm{\omega}$,
(ii) $\bm{\alpha}\cdot\Delta\bm{v}=0$ and $(\bm{\xi}_*^\prime-\bm{\xi}_*)=-(\bm{\xi}^\prime-\bm{\xi}) \parallel \bm{\alpha}$ [see \eqref{eq:2.5}],
and (iii) \eqref{eq:momentum} and \eqref{eq:energy}.
Second, $J^L(f_e)$ is transformed by reversing the direction of $\bm{\alpha}$ as
\begin{equation}
J^L(f_e)=-\frac{\sigma^2}{m}\int g(\bm{X}_{\sigma\bm{\alpha}}^+,\bm{X})
f_{e*}(\bm{X}_{\sigma\bm{\alpha}}^+)f_e(\bm{X})V_\alpha\theta(-V_\alpha)d\Omega(\bm{\alpha})d\bm{\xi}_*.
\end{equation}
\noindent Consequently, $J(f_e)=J^G(f_e)-J^L(f_e)$ is simplified into
\begin{equation}
J(f_e)=\frac{\sigma^2}{m}
\int g(\bm{X}_{\sigma\bm{\alpha}}^+,\bm{X})
f_{e*}(\bm{X}_{\sigma\bm{\alpha}}^+)f_e(\bm{X})V_\alpha
d\Omega(\bm{\alpha})d\bm{\xi}_*.
\end{equation}
\noindent Starting with this form, $J(f_e)$ is further transformed as
\begin{align}
&\frac{\sigma^2}{m}
\int g(\bm{X}_{\sigma\bm{\alpha}}^+,\bm{X})
f_{e*}(\bm{X}_{\sigma\bm{\alpha}}^+)f_e(\bm{X})V_\alpha
d\Omega(\bm{\alpha})d\bm{\xi}_* \Big(=J(f_e)\Big)\notag\displaybreak[0]\\
=&\frac{\sigma^2}{m}f_e(\bm{X})
\int g(\bm{X}_{\sigma\bm{\alpha}}^+,\bm{X})
f_{e*}(\bm{X}_{\sigma\bm{\alpha}}^+)[\bm{c}_*(\bm{X}_{\sigma\bm{\alpha}}^+)-\bm{c}(\bm{X}_{\sigma\bm{\alpha}}^+)]\cdot\bm{\alpha}
d\bm{c}_*(\bm{X}_{\sigma\bm{\alpha}}^+) d\Omega(\bm{\alpha}) \notag\displaybreak[0]\\
=&-\frac{\sigma^2}{m}f_e(\bm{X})
\int g(\bm{X}_{\sigma\bm{\alpha}}^+,\bm{X})
f_{e*}(\bm{X}_{\sigma\bm{\alpha}}^+)\bm{c}(\bm{X}_{\sigma\bm{\alpha}}^+)\cdot\bm{\alpha}
d\bm{c}_*(\bm{X}_{\sigma\bm{\alpha}}^+) d\Omega(\bm{\alpha}) \notag\displaybreak[0]\\
=&-\frac{\sigma^2}{m}f_e(\bm{X})
\int g(\bm{X}_{\sigma\bm{\alpha}}^+,\bm{X})
\rho(\bm{X}_{\sigma\bm{\alpha}}^+)\bm{c}(\bm{X}_{\sigma\bm{\alpha}}^+)\cdot\bm{\alpha}
d\Omega(\bm{\alpha}) \notag\displaybreak[0]\\
=&-\frac{\sigma^2}{m}f_e(\bm{X})
\int g(\bm{X}_{\sigma\bm{\alpha}}^+,\bm{X})
\rho(\bm{X}_{\sigma\bm{\alpha}}^+)\bm{c}(\bm{X})\cdot\bm{\alpha}
d\Omega(\bm{\alpha}) \notag\displaybreak[0]\\
=&-\frac{\sigma^2}{m}f_e(\bm{X})(\bm{\xi}-\bm{v}(\bm{X}))\cdot
\int \bm{\alpha} g(\bm{X}_{\sigma\bm{\alpha}}^+,\bm{X})
\rho(\bm{X}_{\sigma\bm{\alpha}}^+) d\Omega(\bm{\alpha}),\label{eq:Jfe}
\end{align}
\noindent where  $\bm{c}_*(\bm{X})=\bm{\xi}_*-\bm{v}(\bm{X})$,
 $\bm{c}(\bm{X})=\bm{\xi}-\bm{v}(\bm{X})$,
$V_\alpha=(\bm{\xi}_*-\bm{\xi})\cdot\bm{\alpha}$,
and again $\bm{v}(\bm{X}_{\sigma\bm{\alpha}}^+)\cdot\bm{\alpha}=\bm{v}(\bm{X})\cdot\bm{\alpha}$ (or $\Delta\bm{v}\cdot\bm{\alpha}=0$) has been used.
Hence, \eqref{eq:RHS} is obtained.
In the meantime, the left-hand side of \eqref{ES} is transformed as
\begin{align}
 \frac{\partial f_e}{\partial t}
+\xi_i\frac{\partial f_e}{\partial X_i}
=&\Big(\frac{\partial \ln\rho}{\partial t}
+\frac{(\bm{\xi}-\bm{v})}{RT}\cdot\frac{\partial\bm{v}}{\partial t}
+(\frac{(\bm{\xi}-\bm{v})^2}{2RT}-\frac32)\frac{d \ln T}{d t}
 \notag\\
&+v_i\frac{\partial \ln\rho}{\partial X_i}
+v_i\frac{(\bm{\xi}-\bm{v})}{RT}\cdot\frac{\partial\bm{v}}{\partial X_i} \notag \\
&+(\xi_i-v_i)\frac{\partial \ln\rho}{\partial X_i}
+(\xi_i-v_i)\frac{(\bm{\xi}-\bm{v})}{RT}\cdot\frac{\partial\bm{v}}{\partial X_i}\Big)f_e.\label{eq:LHS}
\end{align}
\noindent Comparing \eqref{eq:RHS} and \eqref{eq:LHS}, the following identities are obtained:
\begin{equation}
\frac{\partial \ln\rho}{\partial t}
-\frac32\frac{d \ln T}{d t}
+v_i\frac{\partial \ln\rho}{\partial X_i}=0,\label{eq:const}
\end{equation}
\begin{equation}
 \frac{1}{RT}\frac{\partial{v}_i}{\partial t}
+\frac{v_j}{RT}\frac{\partial{v}_i}{\partial X_j}+\frac{\partial \ln\rho}{\partial X_i}
=-\frac{\sigma^2}{m}\int {\alpha}_i\rho(\bm{X}^+_{\sigma\bm{\alpha}})
   g(\bm{X},\bm{X}^+_{\sigma\bm{\alpha}})d\Omega(\bm{\alpha}),\label{eq:linear}
\end{equation}
\begin{equation}
\frac{d \ln T}{d t}\delta_{ij}
 +\frac{\partial{v}_j}{\partial X_i}+\frac{\partial{v}_i}{\partial X_j}=0.\label{eq:square2}
\end{equation}
\noindent 
{ Here, the time derivative of $T$ is the ordinary derivative because $T$ is independent of $\bm{X}$, as noted immediately after \eqref{eq:Lag}.}
From \eqref{eq:square2}, 
\begin{equation}
\frac32\frac{d \ln T}{d t}
 +\frac{\partial{v}_i}{\partial X_i}=0,\label{eq:square3}
\end{equation}
\noindent holds, and \eqref{eq:const} combined with \eqref{eq:square3} is just the  continuity equation.

In the process from \eqref{eq:LHS} to \eqref{eq:square3}, the specific form of $\bm{v}$,
i.e., \eqref{eq:velocity}, is not fully taken into account.
By using \eqref{eq:velocity}, further simplification is possible.
Indeed,
\begin{equation}
\frac{\partial v_i}{\partial X_j}=\epsilon_{ijk}\omega_k,
\end{equation}
\noindent and thus
\begin{equation}
 \frac{\partial v_i}{\partial X_j}
+\frac{\partial v_j}{\partial X_i}=0,\quad
\frac{\partial v_i}{\partial X_i}=0.
\end{equation}
\noindent Consequently, by \eqref{eq:square3}, 
\begin{equation}
T=\mathrm{const.},\label{eq:Tconst}
\end{equation}
\noindent and \eqref{eq:const} and
\eqref{eq:linear} are reduced to
\begin{align}
 \frac{\partial \ln\rho}{\partial t}
+(u_i+\epsilon_{ijk}X_j\omega_k)\frac{\partial \ln\rho}{\partial X_i}=&0,\label{eq:lnrho}\displaybreak[0]\\
 \frac{d{u}_i}{d t}+\epsilon_{ijk}X_j\frac{d\omega_k}{dt}
+(u_j+\epsilon_{jkl}X_k & \omega_l)\epsilon_{jmi}\omega_m
+ RT\frac{\partial \ln\rho}{\partial X_i} \notag\displaybreak[0]\\
=&-RT\frac{\sigma^2}{m}\int {\alpha}_i\rho(\bm{X}^+_{\sigma\bm{\alpha}})
  g(\bm{X},\bm{X}^+_{\sigma\bm{\alpha}})d\Omega(\bm{\alpha}).
\label{eq:blast}
\end{align}
\noindent Since $\epsilon_{jkl}\epsilon_{jmi}=\delta_{km}\delta_{li}-\delta_{ki}\delta_{lm}$,
the third term of \eqref{eq:blast} is further simplified as 
\begin{align}
(u_j+\epsilon_{jkl}X_k\omega_l)\epsilon_{jmi}\omega_m
=&
\epsilon_{ijk}u_j \omega_k
+\epsilon_{jkl}\epsilon_{jmi} X_k \omega_l \omega_m\notag\displaybreak[0]\\
=&\epsilon_{ijk}u_j \omega_k
+(\delta_{km}\delta_{li}-\delta_{ki}\delta_{lm})X_k \omega_l \omega_m\notag\displaybreak[0]\\
=&\epsilon_{ijk}u_j \omega_k+(X_k \omega_i-X_i \omega_k) \omega_k,
\end{align}
\noindent and \eqref{eq:blast} is finally reduced to
\begin{align}
 \frac{d{u}_i}{d t}+\epsilon_{ijk}u_j \omega_k
+\epsilon_{ijk}X_j\frac{d\omega_k}{dt}
+&\omega_k(X_k\omega_i-X_i\omega_k)
+RT\frac{\partial \ln\rho}{\partial X_i} \notag\displaybreak[0]\\
=&-RT\frac{\sigma^2}{m}\int {\alpha}_i\rho(\bm{X}^+_{\sigma\bm{\alpha}})
  g(\bm{X},\bm{X}^+_{\sigma\bm{\alpha}})d\Omega(\bm{\alpha}).
\label{eq:last}
\end{align}
\noindent The solutions $\rho(t,\bm{X})$, $\bm{u}(t)$, and $\bm{\omega}(t)$ 
for \eqref{eq:lnrho} and \eqref{eq:last}, together with the constant uniform temperature,
determine the local Maxwellian that is admissible as a solution of the Enskog equation.

\begin{remark}
The Boltzmann equation admits a local Maxwellian with radial flow and uniform temperature, both of which may depend on $t$ \cite{G49,S07}.
In this sense, the present result
is more restrictive than the case of the Boltzmann equation.
See also Appendix~\ref{sec:app}.
Although the constant temperature was already pointed out in the seminal work of Resibois \cite{R78}, a rigid body rotation was not brought to attention there.
{Rigid} body rotation was mentioned by Maynar \textit{et al.} \cite{MGB18}, but no details were given.
\end{remark}

Some details of the properties of $g_2$ to be used in Secs.~\ref{sec:ubd} and \ref{sec:bdd} are given in Appendix~\ref{sec:appB}.
Due to the possibility of chained influence of many molecules, 
available properties in the case of the modified Enskog equation are limited,
compared to the Boltzmann--Enskog equation ($g_2\equiv 1$) and 
the original Enskog equation ($g_2$ is a function, not a functional, of density),
 especially for the domain with boundary.
Since the limited properties remain valid for these equations, the results in Secs.~\ref{sec:ubd} and \ref{sec:bdd} are also valid for the Boltzmann--Enskog and original Enskog equations. 

\subsection{Domain without boundary\label{sec:ubd}}
Let us first consider simple situations where there is no physical boundary. 
Because there is no boundary, $\bm{\alpha}$ can take any direction, no matter where $\bm{X}$ is. Moreover, $g$ can be replaced with $g_2$, because $\chi_D(\bm{X})=\chi_D(\bm{X}_{\sigma\bm{\alpha}}^+)\equiv 1$. 
\begin{enumerate}
\item
Suppose that $\rho$ is independent of $\bm{X}$. 
Then, $\rho$ is also independent of $t$ by \eqref{eq:lnrho}
and thus $\rho$ is constant.
In the meantime, $w$ and $Y$ in Appendix~\ref{sec:appB} can be consistently assumed to be independent of $\bm{X}$. We will consider the solution under this assumption. Then, $g_2(\bm{X},\bm{X}_{\sigma\bm{\alpha}}^+)$ does not depend on $\bm{\alpha}$, and thus the integration in \eqref{eq:last} vanishes. 
Consequently, it follows that
\begin{align}
& \frac{d{u}_i}{d t}+\epsilon_{ijk}u_j \omega_k
=0, \label{eq:Xind1}\displaybreak[0]\\
&\epsilon_{ijk}X_j\frac{d\omega_k}{dt}
+\omega_k(X_k\omega_i-X_i\omega_k)=0. \label{eq:Xind2}
\end{align}
\noindent The inner product of \eqref{eq:Xind2} and $\bm{X}$ shows that $\bm{X}\parallel\bm{\omega}$. Since $\bm{\omega}$ is independent of $\bm{X}$, $\bm{\omega}$ should be zero, and accordingly $\bm{u}$ is a constant vector by \eqref{eq:Xind1}. This is a time-independent uniform state with a constant flow velocity. 
\item
Axisymmetric solution:
Introduce the cylindrical coordinates $(P,\phi,z)$ for $\bm{X}$ and 
corresponding unit basis vectors $(\bm{e}_P,\bm{e}_\phi,\bm{e}_z)$. 
Let $\alpha_P$, $\alpha_\phi$, and $\alpha_z$ be the components of $\bm{\alpha}$
in the directions of $\bm{e}_P$, $\bm{e}_\phi$, and $\bm{e}_z$, respectively:
$\bm{\alpha}=\alpha_P\bm{e}_P+\alpha_\phi\bm{e}_\phi+\alpha_z\bm{e}_z$.
Now assume that the state is independent of $\phi$. In this case, $\partial/\partial\phi=0$ and an admissible flow velocity is restricted to the form $\bm{u}=u\bm{e}_z$, $\bm{\omega}=\omega\bm{e}_z$, i.e., $\bm{v}=u\bm{e}_z-P\omega\bm{e}_\phi$. Then, the equations \eqref{eq:lnrho} and \eqref{eq:last} are reduced to 
\begin{subequations}\begin{align}
& \frac{\partial \ln\rho}{\partial t}
+ u\frac{\partial \ln\rho}{\partial z}=0,\label{eq:cyl1}\displaybreak[0]\\
& \frac{d{u}}{d t}
+RT\frac{\partial \ln\rho}{\partial z}
=-RT\frac{\sigma^2}{m}\int {\alpha}_z\rho(\bm{X}^+_{\sigma\bm{\alpha}})
  g_2(\bm{X},\bm{X}^+_{\sigma\bm{\alpha}})d\Omega(\bm{\alpha}),
\label{eq:cyl2}\displaybreak[0]\\
&-P\frac{d\omega}{dt}
=-RT\frac{\sigma^2}{m}\int {\alpha}_\phi\rho(\bm{X}^+_{\sigma\bm{\alpha}})
  g_2(\bm{X},\bm{X}^+_{\sigma\bm{\alpha}})d\Omega(\bm{\alpha}),
\label{eq:cyl3}\displaybreak[0]\\
& 
-P\omega^2+RT\frac{\partial \ln\rho}{\partial P}
=-RT\frac{\sigma^2}{m}\int {\alpha}_P\rho(\bm{X}^+_{\sigma\bm{\alpha}})
  g_2(\bm{X},\bm{X}^+_{\sigma\bm{\alpha}})d\Omega(\bm{\alpha}).\label{eq:cyl4}
\end{align}\end{subequations}
\noindent Before proceeding, it should be noted that
the distance $P^\prime$ of the position $\bm{X}_{\sigma\bm{\alpha}}^+$ from the central axis 
can be expressed as $P^\prime=(P^2+\sigma^2\sin^2\theta_\alpha+2P\sigma\sin\theta_\alpha\cos\varphi_\alpha)^{1/2}$, where $(\theta_\alpha,\varphi_\alpha)$ is a pair of the polar and azimuthal angles of $\bm{\alpha}$ with $\bm{e}_z$ being the polar direction, i.e., $\alpha_P=\sin\theta_\alpha\cos\varphi_\alpha$, $\alpha_\phi=\sin\theta_\alpha\sin\varphi_\alpha$, and $\alpha_z=\cos\theta_\alpha$.
Moreover, because of \eqref{eq:g2even} in Appendix~\ref{sec:appB}, $g_2(\bm{X},\bm{X}_{\sigma\bm{\alpha}}^+)=g_2(\bm{X},\bm{X}_{\sigma\bm{\beta}}^+)$ holds for $\bm{\beta}\equiv \bm{\alpha}-2\alpha_\phi\bm{e}_\phi$. 
Hence, $g_2$ is even in $\varphi_\alpha$ (or $\alpha_\phi$).
Since $\rho(\bm{X}_{\sigma\bm{\alpha}}^+)$ is a function of $P^\prime$ and $z+\sigma\alpha_z$,
it is also even in  $\varphi_\alpha$.
Therefore, the integrand of \eqref{eq:cyl3} is odd with respect to $\varphi_\alpha$, and
the right-hand side of \eqref{eq:cyl3} vanishes by the integration with respect to $\varphi_\alpha$, yielding that $\omega=\mathrm{const}$.
\begin{enumerate}
\item[a.]
Suppose that $\rho$ is independent of $P$.
Then, $w$ and $Y$ in Appendix~\ref{sec:appB} can be consistently assumed to have the same property. We will consider the solution under this assumption. Then, $g_2(\bm{X}_1,\bm{X}_2)$ is a function of $z_1$, $z_2$, and $|\bm{X}_1-\bm{X}_2|$ only, where $z_1$ and $z_2$ are the $z$-coordinates of $\bm{X}_1$ and $\bm{X}_2$, respectively; see \eqref{eq:g2} in Appendix~\ref{sec:appB}.
{Recall} that the $z$-coordinate of $\bm{X}_{\sigma\bm{\alpha}}^+$ is given by
$z+\sigma\alpha_z=z+\sigma\cos\theta_\alpha$.
The integrand of \eqref{eq:cyl4} is thus simply proportional to $\cos\varphi_\alpha$ through $\alpha_P$,
since both $\rho(\bm{X}_{\sigma\bm{\alpha}}^+)$ and $g(\bm{X},\bm{X}_{\sigma\bm{\alpha}}^+)$ are independent of $\varphi_\alpha$.
Consequently, the integral in \eqref{eq:cyl4} vanishes by the integration with respect to $\varphi_\alpha$. Hence, $\omega=0$, and $\rho$ and $u$ are determined by \eqref{eq:cyl1} and \eqref{eq:cyl2}:
\begin{align}
& \frac{\partial \ln\rho}{\partial t}
+ u\frac{\partial \ln\rho}{\partial z}=0,\displaybreak[0]\\
& \frac{d{u}}{d t}
+RT\frac{\partial \ln\rho}{\partial z}
=-RT\frac{\sigma^2}{m}\int {\alpha}_z\rho(\bm{X}^+_{\sigma\bm{\alpha}})
  g_2(\bm{X},\bm{X}^+_{\sigma\bm{\alpha}})d\Omega(\bm{\alpha}).
\end{align}
\noindent This is a uniform flow along the axial direction.
\item[b.]
Suppose that $\rho$ is independent of $z$, and thus the system is invariant under a translation in the $z$-direction. Then, $\rho$ is independent of $t$ as well by \eqref{eq:cyl1}. 
Moreover, $g(\bm{X},\bm{X}_{\sigma\bm{\alpha}}^+)$ is even with respect to $\alpha_z$
by \eqref{eq:g2zeven} in Appendix~\ref{sec:appB}. 
Since $P^\prime=(P^2+\sigma^2\sin^2\theta_\alpha+2P\sigma\sin\theta_\alpha\cos\varphi_\alpha)^{1/2}$, the integrand in \eqref{eq:cyl2} is odd in $\alpha_z=\cos\theta_\alpha$
and the right-hand side of \eqref{eq:cyl2} vanishes by the integration with respect to $\theta_\alpha$. Therefore, $u$ is constant and $\rho$ is a function of $P$ determined by \eqref{eq:cyl4}:
\begin{equation}\label{eq:ubd_tube}
-P\frac{\omega^2}{RT}+\frac{d \ln\rho}{d P}
=-\frac{\sigma^2}{m}\int {\alpha}_P\rho(P^\prime)
  g_2(\bm{X},\bm{X}^+_{\sigma\bm{\alpha}})d\Omega(\bm{\alpha}).
\end{equation}
\noindent This is a superposition of a time-independent rigid body rotation and a constant uniform flow along the axis.
\item[c.]
Suppose $u$ is zero. Then, $\rho$ is independent of $t$ and is determined as a function of $P$ and $z$ by \eqref{eq:cyl2} and \eqref{eq:cyl4}:
\begin{align}
& \frac{\partial \ln\rho}{\partial z}
=-\frac{\sigma^2}{m}\int {\alpha}_z\rho(\bm{X}^+_{\sigma\bm{\alpha}})
  g_2(\bm{X},\bm{X}^+_{\sigma\bm{\alpha}})d\Omega(\bm{\alpha}),\displaybreak[0]
\\
& 
-P\frac{\omega^2}{RT}+\frac{\partial \ln\rho}{\partial P}
=-\frac{\sigma^2}{m}\int {\alpha}_P\rho(\bm{X}^+_{\sigma\bm{\alpha}})
  g_2(\bm{X},\bm{X}^+_{\sigma\bm{\alpha}})d\Omega(\bm{\alpha}).
\end{align}
\noindent This is a time-independent rigid body rotation.
\item[d.]
Suppose that $\rho$ is independent of $t$. Then $u=0$, or otherwise $\rho$ is independent of $z$. The case $u=0$ is the same as Case 2c, while the case $u\ne0$ is the same as Case 2b.
\end{enumerate}
\end{enumerate}

\subsection{Domain with boundary\label{sec:bdd}}
Next consider simple situations in a domain with boundary. A few remarks should be made before proceeding. First, we impose
on the boundary $\partial D$ only the \emph{impermeable} condition, i.e., $\bm{v}\cdot\bm{n}=0$, where $\bm{n}$ is the inward unit vector normal to the boundary. 
Second, the range of integration with respect to $\bm{\alpha}$
may be limited at positions near the boundary, although there is no such a limitation away from the boundary. No limitation in the latter implies the assumption that there is a subdomain of $D$ such that $\bm{X}_{\sigma\bm{\alpha}}^+\in D$ for any direction of $\bm{\alpha}$.
\begin{enumerate}
\item
Axisymmetric solution in a circular cylinder:%
\footnote{Note the difference between the boundary $\partial D$ and the surface of the cylinder. The surface is placed outside the boundary $\partial D$ by a distance of $\sigma/2$ from the central axis.}
Since $\partial/\partial\phi=0$, an admissible flow velocity is restricted to the form $\bm{u}=u\bm{e}_z$, $\bm{\omega}=\omega\bm{e}_z$, i.e., $\bm{v}=u\bm{e}_z-P\omega\bm{e}_\phi$, as is already noted in Sec.~\ref{sec:ubd}. This property is not affected by the presence of boundary.
Again, the equations \eqref{eq:lnrho} and \eqref{eq:last} are reduced to 
\begin{subequations}\begin{align}
& \frac{\partial \ln\rho}{\partial t}
+ u\frac{\partial \ln\rho}{\partial z}=0,\label{eq:tube1}\displaybreak[0]\\
& \frac{d{u}}{d t}
+RT\frac{\partial \ln\rho}{\partial z}
=-RT\frac{\sigma^2}{m}\int {\alpha}_z\rho(\bm{X}^+_{\sigma\bm{\alpha}})
  g(\bm{X},\bm{X}^+_{\sigma\bm{\alpha}})d\Omega(\bm{\alpha}),
\label{eq:tube2}\displaybreak[0]\\
&-P\frac{d\omega}{dt}
=-RT\frac{\sigma^2}{m}\int {\alpha}_\phi\rho(\bm{X}^+_{\sigma\bm{\alpha}})
  g(\bm{X},\bm{X}^+_{\sigma\bm{\alpha}})d\Omega(\bm{\alpha}),
\label{eq:tube3}\displaybreak[0]\\
& 
-P\omega^2+RT\frac{\partial \ln\rho}{\partial P}
=-RT\frac{\sigma^2}{m}\int {\alpha}_P\rho(\bm{X}^+_{\sigma\bm{\alpha}})
  g(\bm{X},\bm{X}^+_{\sigma\bm{\alpha}})d\Omega(\bm{\alpha}),
\label{eq:tube4}
\end{align}\end{subequations}
\noindent where the same notation of coordinates as that in Case 2 of Sec.~\ref{sec:ubd} has been used.
Recall that $g(\bm{X},\bm{X}^+_{\sigma\bm{\alpha}})=0$ for $\bm{X}^+_{\sigma\bm{\alpha}}\notin D$. Since the cross-section perpendicular to the axis is circular,
the integration range for $\varphi_\alpha$ is symmetric with respect to $\varphi_\alpha=0$.
Then, because of the similar reason to the case \eqref{eq:cyl3} in Sec.~\ref{sec:ubd}, the integral in \eqref{eq:tube3} vanishes and $\omega=\mathrm{const}$.
\begin{enumerate}
\item[a.] The density $\rho$ cannot be uniform in $P$.
Suppose that $\rho$ is independent of $P$.
Then, \eqref{eq:tube4} is reduced to
\begin{equation}
-P\omega^2
=-RT\frac{\sigma^2}{m}\int {\alpha}_P\rho(\bm{X}^+_{\sigma\bm{\alpha}})
  g(\bm{X},\bm{X}^+_{\sigma\bm{\alpha}})d\Omega(\bm{\alpha}),
\end{equation}
and the left-hand side is not positive.
However, 
since $\alpha_P<0$ and $\rho g>0$ on the boundary $\partial D$,
the right-hand side is positive, which is inconsistent with the left-hand side.
\item[b.]
Suppose that $\rho$ is independent of $z$, and thus the system is invariant under a translation in the $z$-direction. Then, $\rho$ is independent of $t$ as well by \eqref{eq:tube1}.
Consequently, the right-hand side of \eqref{eq:tube2} is time-independent and $du/dt$ has to be constant. Meanwhile, because of the similar reason to that in Case 2b of Sec.~\ref{sec:ubd},
the right-hand side of \eqref{eq:tube2} vanishes and $u$ is constant. 
Therefore, $\rho$ is determined as a function of $P$ by \eqref{eq:tube4}:
\begin{equation}\label{eq:bff_tube}
-P\frac{\omega^2}{RT}+\frac{d \ln\rho}{d P}
=-\frac{\sigma^2}{m}\int {\alpha}_P\rho(P^\prime)
  g(\bm{X},\bm{X}^+_{\sigma\bm{\alpha}})d\Omega(\bm{\alpha}).
\end{equation}
\noindent This is a time-independent rigid body rotation superposed with a constant uniform flow along the axis. 
\item[c.]
Suppose that $u$ is zero. Then, $\rho$ is independent of $t$ by \eqref{eq:tube1} and is determined as a function of $P$ and $z$ by \eqref{eq:tube2} and \eqref{eq:tube4}:
\begin{align}
& \frac{\partial \ln\rho}{\partial z}
=-\frac{\sigma^2}{m}\int {\alpha}_z\rho(\bm{X}^+_{\sigma\bm{\alpha}})
  g(\bm{X},\bm{X}^+_{\sigma\bm{\alpha}})d\Omega(\bm{\alpha}),\displaybreak[0]
\\
& 
-P\frac{\omega^2}{RT}+\frac{\partial \ln\rho}{\partial P}
=-\frac{\sigma^2}{m}\int {\alpha}_P\rho(\bm{X}^+_{\sigma\bm{\alpha}})
  g(\bm{X},\bm{X}^+_{\sigma\bm{\alpha}})d\Omega(\bm{\alpha}).
\end{align}
\noindent This is a time-independent rigid body rotation.
\end{enumerate}
\item 
Axisymmetric solution in a sphere:%
\footnote{Note the difference between the boundary $\partial D$ and the surface of the sphere. The surface is placed outside the boundary $\partial D$ by a distance of $\sigma/2$ from the center.}
It is convenient to introduce the spherical coordinates $(r,\theta,\varphi)$ for $\bm{X}$ and corresponding unit basis vectors $(\bm{e}_r, \bm{e}_\theta, \bm{e}_\varphi)$.
Let $\alpha_r$, $\alpha_\theta$, and $\alpha_\varphi$ be the components of $\bm{\alpha}$
in the directions of $\bm{e}_r$, $\bm{e}_\theta$, and $\bm{e}_\varphi$, respectively:
$\bm{\alpha}=\alpha_r\bm{e}_r+\alpha_\theta\bm{e}_\theta+\alpha_\varphi\bm{e}_\varphi$.
Now assume that the state is independent of $\varphi$. In this case, $\partial/\partial\varphi=0$
and the flow velocity is compatible with the axisymmetric condition
in the form $\bm{u}=u\bm{e}_z$ and $\bm{\omega}=\omega\bm{e}_z$.
Note that $\bm{X}=r\bm{e}_r$ and $\bm{e}_z=\bm{e}_r\cos\theta-\bm{e}_\theta\sin\theta$.
$\bm{v}=u\cos\theta\bm{e}_r-u\sin\theta\bm{e}_\theta-r\omega\sin\theta\bm{e}_\varphi$.
Then, \eqref{eq:lnrho} and \eqref{eq:last} are reduced to 
\begin{align}
 \frac{\partial \ln\rho}{\partial t}
+&u\cos\theta\frac{\partial \ln\rho}{\partial r}
-\frac{u\sin\theta}{r}\frac{\partial \ln\rho}{\partial\theta}=0,\label{eq:sp1}\displaybreak[0]\\
\frac{d{u}}{d t}\cos\theta&
-r \omega^2\sin^2\theta
+RT\frac{\partial \ln\rho}{\partial r}\notag\\
=&-RT\frac{\sigma^2}{m}\int {\alpha}_r\rho(\bm{X}^+_{\sigma\bm{\alpha}})
  g(\bm{X},\bm{X}^+_{\sigma\bm{\alpha}})d\Omega(\bm{\alpha}),\label{eq:sp2}\displaybreak[0]\\
-\frac{d{u}}{d t}\sin\theta&
-r\omega^2\cos\theta\sin\theta
+\frac{RT}{r}\frac{\partial \ln\rho}{\partial \theta} \notag\\
=&-RT\frac{\sigma^2}{m}\int {\alpha}_\theta\rho(\bm{X}^+_{\sigma\bm{\alpha}})
  g(\bm{X},\bm{X}^+_{\sigma\bm{\alpha}})d\Omega(\bm{\alpha}),\label{eq:sp3}\displaybreak[0]\\
-r\frac{d\omega}{dt}\sin\theta&
=-RT\frac{\sigma^2}{m}\int {\alpha}_\varphi\rho(\bm{X}^+_{\sigma\bm{\alpha}})
  g(\bm{X},\bm{X}^+_{\sigma\bm{\alpha}})d\Omega(\bm{\alpha}).\label{eq:sp4}
\end{align}
\noindent Before proceeding, let $(r^\prime,\theta^\prime,\varphi^\prime)$ be the spherical coordinates of $\bm{X}_{\sigma\bm{\alpha}}^+$ 
 and let $(\theta_\alpha,\varphi_\alpha)$ be the polar and azimuthal angles of $\bm{\alpha}$
 with $\bm{e}_r$ being the polar direction:
\begin{align}
&
\bm{X}_{\sigma\bm{\alpha}}^+\equiv \bm{X}+\sigma\bm{\alpha}=(r+\sigma\alpha_r)\bm{e}_r+\sigma\alpha_\theta\bm{e}_\theta+\sigma\alpha_\varphi\bm{e}_\varphi,\displaybreak[0]\\
&
\alpha_r=\cos\theta_\alpha,\quad 
\alpha_\theta=\sin\theta_\alpha\cos\varphi_\alpha,\quad 
\alpha_\varphi=\sin\theta_\alpha\sin\varphi_\alpha,
\displaybreak[0]\\
&
\bm{e}_{r}\cdot\bm{e}_z=\cos\theta,\quad
\bm{e}_{\theta}\cdot\bm{e}_z=-\sin\theta,\quad
\bm{e}_{\varphi}\cdot\bm{e}_z=0,\displaybreak[0]\\
&
\bm{e}_{r}\cdot\bm{e}_x=\sin\theta\cos\varphi,\quad
\bm{e}_{\theta}\cdot\bm{e}_x=\cos\theta\cos\varphi,\quad
\bm{e}_{\varphi}\cdot\bm{e}_x=-\sin\varphi,
\displaybreak[0]\\
&
r^\prime=\sqrt{|\bm{X}+\sigma\bm{\alpha}|^2}=\sqrt{r^2+\sigma^2+2r\sigma\cos\theta_\alpha},
\displaybreak[0]\\
&r^\prime\cos\theta^\prime
=\bm{X}_{\sigma\bm{\alpha}}^+\cdot\bm{e}_z
=(r+\sigma\cos\theta_\alpha)\cos\theta-\sigma\sin\theta_\alpha\cos\varphi_\alpha\sin\theta.
%
\end{align}
\noindent Obviously $\theta^\prime$ depends on $\varphi_\alpha$ as a function of $\cos\varphi_\alpha$, while $r^\prime$ is independent of $\varphi_\alpha$. 
Because the system is axisymmetric, \eqref{eq:g2even} in Appendix~\ref{sec:appB} applies, and  
$g(\bm{X},\bm{X}_{\sigma\bm{\alpha}}^+)=g(\bm{X},\bm{X}_{\sigma\bm{\beta}}^+)$ holds for $\bm{\beta}=\bm{\alpha}-2\alpha_\varphi\bm{e}_\varphi$. Therefore
$g(\bm{X},\bm{X}_{\sigma\bm{\alpha}}^+)$ is even in $\varphi_\alpha$.
Since the integration is over the whole range of $\varphi_\alpha$, 
the integral in \eqref{eq:sp4} becomes zero, yielding that $\omega$ is constant.
Furthermore, $u\equiv0$, since the boundary is impermeable.
Hence, $\rho$ is independent of $t$ by \eqref{eq:sp1},
and \eqref{eq:sp2} and \eqref{eq:sp3} are reduced to 
\begin{align}
&
-\frac{r \omega^2}{RT}\sin^2\theta
+\frac{\partial \ln\rho}{\partial r}
=-\frac{\sigma^2}{m}\int {\alpha}_r\rho(\bm{X}^+_{\sigma\bm{\alpha}})
  g(\bm{X},\bm{X}^+_{\sigma\bm{\alpha}})d\Omega(\bm{\alpha}),\label{eq:sp1a}\displaybreak[0]\\
&
-\frac{r \omega^2}{RT}\cos\theta\sin\theta
+\frac{1}{r}\frac{\partial \ln\rho}{\partial \theta}
=-\frac{\sigma^2}{m}\int {\alpha}_\theta\rho(\bm{X}^+_{\sigma\bm{\alpha}})
  g(\bm{X},\bm{X}^+_{\sigma\bm{\alpha}})d\Omega(\bm{\alpha}).
\label{eq:sp2a}
\end{align}
\begin{enumerate}
\item[a.]
The density $\rho$ cannot be independent of $r$.
Suppose that $\rho$ is independent of $r$. Then, \eqref{eq:sp1a} is reduced to
\begin{equation}
-\frac{r \omega^2}{RT}\sin^2\theta
=-\frac{\sigma^2}{m}\int {\alpha}_r\rho(\bm{X}^+_{\sigma\bm{\alpha}})
  g(\bm{X},\bm{X}^+_{\sigma\bm{\alpha}})d\Omega(\bm{\alpha}),
\end{equation}
and thus, the left-hand side is not positive.
Meanwhile, since $\rho g>0$ and $\alpha_r<0$ on the boundary,
the right-hand side is positive and is inconsistent with the left-hand side.
\item[b.]
Suppose that $\rho$ is independent of $\theta$.
Then, $w$ and $Y$ in Appendix~\ref{sec:appB} can be consistently assumed to be spherically symmetric. We will consider the solution under this assumption.
Then, $g_2(\bm{X}_1,\bm{X}_2)$ is a function of $r_1$, $r_2$, and $\bm{X}_1\cdot\bm{X}_2$ only, where $r_1$ and $r_2$ are the radial coordinates of $\bm{X}_1$ and $\bm{X}_2$, respectively, see \eqref{eq:g2} in Appendix~\ref{sec:appB}.
Since $\bm{X}\cdot\bm{X}_{\sigma\bm{\alpha}}^+=(r+\sigma\alpha_r)r$,
$g_2$ and $g$ are independent of $\varphi_\alpha$. 
Consequently, the integral in \eqref{eq:sp2a} vanishes
by the integration with respect to $\varphi_\alpha$.
Hence $\omega=0$, and $\rho$ is determined as a function of $r$ by
\begin{equation}
  \frac{d\ln\rho}{dr}
=-\frac{\sigma^2}{m}\int {\alpha}_r\rho(\bm{X}^+_{\sigma\bm{\alpha}})
  g(\bm{X},\bm{X}^+_{\sigma\bm{\alpha}})d\Omega(\bm{\alpha}).
\end{equation}
This is a spherically symmetric time-independent resting state.
\end{enumerate}
\end{enumerate}

\section{Numerical examples\label{sec:ND}}

\begin{figure}\centering
\noindent\begin{tabular}{ccc}
\includegraphics[width=0.475\textwidth]{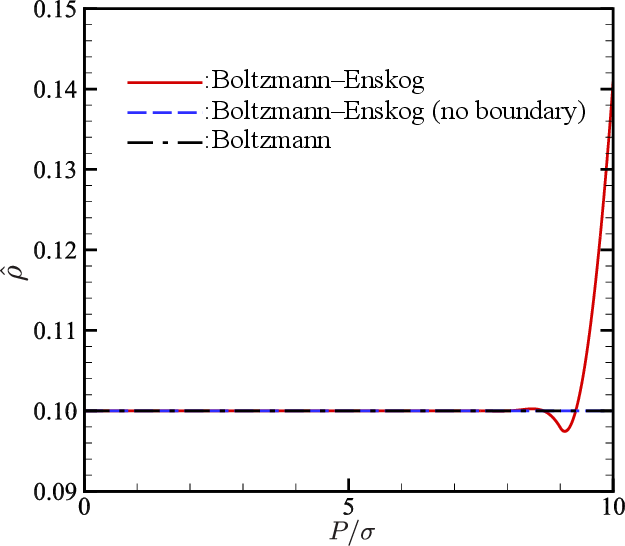} &
\includegraphics[width=0.475\textwidth]{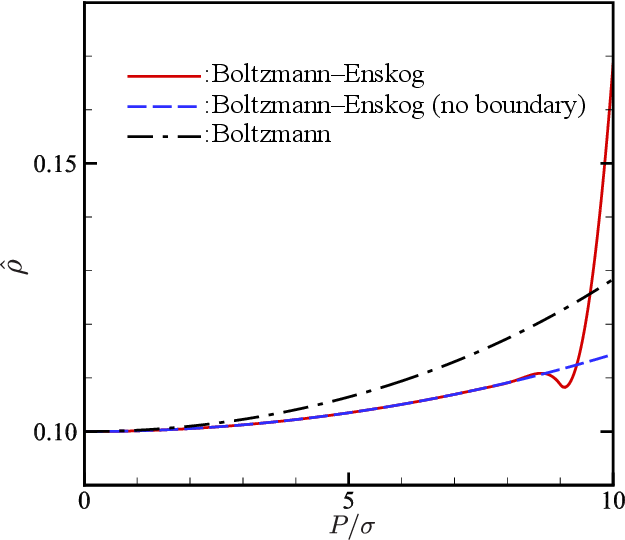} \\
(a) & (b)
\end{tabular}
\caption{\label{fig:1}Density profile for axisymmetric solutions, i.e., Case 2b in Sec.~\ref{sec:ubd} and Case 1b in Sec.~\ref{sec:bdd}. (a) $\omega=0$ (no rotation), (b) $\omega=0.05\sqrt{2RT}/\sigma$. 
The {$\hat{\rho}$ is a} normalized density {defined by} $\hat{\rho}\equiv(\pi/6)(\sigma^3/m)\rho${, which} 
represents the local volume fraction of molecules { and}  never exceed {the value of close-packing of equal spheres $\sqrt{2}\pi/6\simeq 0.74$ in the case of the Boltzmann--Enskog equation}.
In both panels, solid (red) lines indicate the results of the Boltzmann--Enskog equation in a circular cylinder with a radius of $10.5\sigma$ (Case 1b in Sec.~\ref{sec:bdd}), dashed (blue) lines those of the Boltzmann--Enskog equation without boundary (Case 2b in Sec.~\ref{sec:ubd}), and dash-dotted (black) lines those of the Boltzmann equation
{ (the solution of \eqref{eq:ubd_tube} or \eqref{eq:bff_tube} with the integral on the } { right-hand side being omitted)}.}
\end{figure}

We present numerical examples for the Boltzmann--Enskog equation, i.e., $g_2=1$. 
Case 2b in Sec.~\ref{sec:ubd} and Case 1b in Sec.~\ref{sec:bdd} are chosen as the simplest examples. 
It should be reminded that we simply impose the condition $\bm{v}\cdot\bm{n}=0$
on the boundary, see the first paragraph of Sec.~\ref{sec:bdd}.
Figure~\ref{fig:1} shows the axisymmetric solution with and without rotation. 
In Fig.~\ref{fig:1}\textbf{a}, since there is no rotation, the Boltzmann--Enskog equation gives the uniform density in the case without boundary as does the Boltzmann equation. 
However, the density profile is no longer uniform in the case with boundary. 
Figure~\ref{fig:1}\textbf{b} shows the density profile in the case of a rigid body rotation. 
In the case without boundary, the Boltzmann--Enskog equation gives a monotonically increasing density with the distance from the axis of rotation, as does the Boltzmann equation. 
Further numerical experiments by varying the computational domain show an unlimited increase in density, although the rate of increase is smaller than the case of the Boltzmann equation.
Indeed, the behavior of density at a far distance can be estimated
by retaining the first two terms of the Taylor expansion of $\rho(P^\prime)$ 
 around $P$ in \eqref{eq:ubd_tube}:
$\rho(P^\prime)\simeq\rho(P)+(1/2)\sigma\sin\theta_\alpha(2\cos\varphi_\alpha+\varepsilon\sin\theta_\alpha\sin^2\varphi_\alpha)(d\rho/dP)$, where $\varepsilon=\sigma/P$ and its higher order terms have been neglected. Using this approximation leads to the following expression%
\footnote{We have the same expression as \eqref{eq:LW} for the entire region, without approximation, from the compressible Navier--Stokes--Fourier set of equations, with the aid of the equation of state $p=\rho RT[1+(2\pi/3)(\sigma^3/m)\rho]$. This equation of state is that for the Boltzmann--Enskog equation, see ,e.g., \cite{CC95,HCB64}. In the rigid body rotation mode, the viscous dissipation into heat does not occur and the isothermal state is compatible with the energy equation.\label{fn:4}}:
\begin{equation}
\rho(P)\simeq C \exp(-W_0(\frac{4\pi}{3}\frac{\sigma^3}{m}C\exp(\frac{\omega^2P^2}{2RT}))+\frac{\omega^2P^2}{2RT}),\label{eq:LW} 
\end{equation}
where $C$ is a positive constant {and
$W_0(x)$ is the principal branch of the Lambert W function \cite{CGHJK96,J14}}.
Since $W_0(x)\approx \ln(x)-\ln(\ln(x))+\cdots$ as $x\to\infty$, 
$\rho(P)\approx [{C} \omega^2/(2RT)]P^2$ as $P\to\infty$.
The unlimited increase of density in the infinite domain is one { of the reasons} why the rigid body rotation mode escaped from the discussions in \cite{R78}.
In the presence of a boundary, the density remains finite, and its profile is no longer monotonic and exhibits the behavior similar to the no-rotation case near the boundary.

{
Although the present numerical study is limited to the Boltzmann--Enskog equation,
some comments on the original and modified Enskog equations are in order.
The non-monotonic profile of density near the boundary is expected 
for these equations as well. However, the growing rate of density is different because of the difference of the equation of state, see the footnote~\ref{fn:4}.
Since the H theorem is not assured,
the numerical study of the original Enskog equation was not carried out in the present work.
Numerical study of the modified Enskog equation is desired,
but remains difficult and untouched.
}

\section{Conclusion\label{sec:conclusion}}
In the present paper, we have discussed the summational invariant and the corresponding local Maxwellian that are compatible with the Enskog equation. Unlike the Boltzmann equation, a general form of the local Maxwellian is not obtained analytically. However, the admissible local Maxwellian turns out to be more restrictive than the case of the Boltzmann equation in the sense that (i) the temperature does not depend on spatial variables nor on time and that (ii) the flow is a superposition of a spatially uniform flow and a rigid body rotation. A radial flow and a time-dependent temperature are not possible, unlike the case of the Boltzmann equation. The influence of a boundary on the admissible local Maxwellian has also been discussed in simple situations; a uniform density profile is no longer established in the presence of a boundary, as is widely recognized.

The possibility of a rigid body rotation was not brought to attention in the seminal work of Resibois \cite{R78}. This is probably due to the fact that the density grows indefinitely in the infinite domain and that the Fourier analysis has been applied to the spatial variables in \cite{R78}. The infinite growth of the density in the infinite domain is confirmed in the present work by both numerical experiments and a far-field estimate. The numerical experiments also demonstrate that a rigid body rotation mode with a finite local density (or more strongly with a local volume fraction less than unity) is possible in an axially symmetric confinement. The rigid body rotation shown in Fig.~\ref{fig:1} is compatible with a specular reflection wall and with other conventional types of wall, such as the diffuse reflection and the Cercignani--Lampis condition. Apart from the specular reflection wall, the wall temperature must be uniform and the wall must rotate about the central axis at the angular velocity $\omega$ (and must move in the axial direction at the velocity $u$).

\appendix

\section{Another approach to the admissible local Maxwellian\label{sec:app}}
In Sec.~\ref{sec:SI}, we have used the conservation of the angular momentum,
in addition to other kinds of conservation used in the case of the Boltzmann equation.
In this Appendix, we will show that the same form of the Maxwellian as in  \eqref{eq:LocalMax} can be obtained without using the angular momentum,
thereby making clearer the origin of the difference with the case of the Boltzmann equation.

Consider the variational problem of \eqref{eq:S1} with respect to twelve variables of molecular velocities under the constraints \eqref{eq:momentum} and \eqref{eq:energy}. Then we recover \eqref{eq:variation} with $\bm{\gamma}=\bm{0}$,
where $\bm{\lambda}$ and $\mu$ are independent of the molecular velocity variables.
Hence, at this stage, we obtain
\begin{equation}
\ln f(\bm{X})=\bm{\lambda}(\bm{X})\cdot\bm{\xi}+\mu(\bm{X})\bm{\xi}^2+\beta(\bm{X}).
\end{equation}
Substitution of the above into \eqref{eq:S1} shows that $\mu$ is independent of $\bm{X}$,
while $\bm{\lambda}(\bm{X})$ needs to satisfy
\begin{equation}
[\bm{\lambda}(\bm{X})-\bm{\lambda}(\bm{X}^-_{\sigma\bm{\alpha}})]\cdot(\bm{\xi}_*-\bm{\xi}_*^\prime)=0.
\end{equation}
Consequently, the form of \eqref{eq:Max} is recovered with a new restriction
\begin{equation}
\Delta\bm{v}\cdot(\bm{\xi}_*-\bm{\xi}_*^\prime)=0,
\quad \mbox{or equivalently}\quad\Delta\bm{v}\cdot\bm{\alpha}=0,\label{eq:app}
\end{equation}
where $\Delta\bm{v}\equiv \bm{v}(\bm{X}_{\sigma\bm{\alpha}}^+)-\bm{v}(\bm{X})$.
Thanks to \eqref{eq:app}, the process of deriving \eqref{eq:RHS} is unchanged
and \eqref{eq:const}--\eqref{eq:square2} are recovered as they stand.
Taking a partial derivative of \eqref{eq:square2} with respect to $\bm{X}$, 
it is seen \cite{G49,S07} that $\bm{v}$ can be written as $v_i(t,\bm{X})=u_i(t)+M_{ij}(t)X_j$.
Thus $\Delta v_i=\sigma M_{ij}(t)\alpha_j$
and accordingly $M_{ij}\alpha_i\alpha_j=0$ by \eqref{eq:app}.
Furthermore, the substitution of the form of $v_i(t,\bm{X})$ into \eqref{eq:square2} gives the relation
$M_{ij}+M_{ji}=-(d\ln T/d t)\delta_{ij}$.
This means that $M_{ij}$ can be expressed as 
$M_{ij}(t)=-(1/2)(d\ln T/d t)\delta_{ij}+\Omega_{ij}(t)$
with $\Omega_{ij}$ being an antisymmetric matrix, i.e., $\Omega_{ij}+\Omega_{ji}=0$.
Finally, substituting the form of $M_{ij}$ into $M_{ij}\alpha_i\alpha_j=0$ yields
\begin{align}
0= M_{ij}\alpha_j\alpha_i
&=-\frac12\frac{d\ln T}{d t}+\Omega_{ij}\alpha_j\alpha_i \notag\\
&=-\frac12\frac{d\ln T}{d t}+\frac12(\Omega_{ij}+\Omega_{ji})\alpha_j\alpha_i 
=-\frac12\frac{d\ln T}{d t}.
\end{align}
Hence $T$ is a constant and $v_i(t,\bm{X})=u_i(t)+\Omega_{ij}(t)X_j$,
the same conclusion as \eqref{eq:Tconst} and \eqref{eq:velocity}.

As is clear from the above discussion, \eqref{eq:app} is the property that restricts
the local Maxwellian to be a superposition of a uniform flow and a rigid body rotation with a constant temperature.
In the discussions in Sec.~\ref{sec:SI}, the property \eqref{eq:app} was embedded as the conservation of the angular momentum.

\section{Some properties of $g_2$ and related quantities\label{sec:appB}}
The purpose of this Appendix is to explain the properties of $g_2$ used in Secs.~\ref{sec:ubd} and \ref{sec:bdd}.

In the framework of the modified Enskog equation,
the velocity distribution function $f$ is assumed to be in the form:
\begin{align}
  f(t,\bm{X}_{1},\bm{\xi}_{1})
=&{ \frac{mN}{\Phi(t)}W(t,\bm{X}_{1},\bm{\xi}_{1}) Y(t,\bm{X}_1),}\label{eq:factorize}
\end{align}
where $N$ is the number of molecules in $D$,
\begin{subequations}\label{eq:partition}\begin{align}
& {Y(t,\bm{X}_1)=\int_{D^{N-1}}w(t,\bm{X}_{2})\cdots w(t,\bm{X}_{N})\Theta(\bm{X}_{1},\cdots,\bm{X}_{N})d\bm{X}_{2}\cdots d\bm{X}_{N},}\label{eq:A.1}\\
 & \Phi(t)=\int_{D^{N}}w(t,\bm{X}_{1})\cdots w(t,\bm{X}_{N})\Theta(\bm{X}_{1},\cdots,\bm{X}_{N})d\bm{X}_{1}\cdots d\bm{X}_{N},\displaybreak[0]\label{eq:A.2}\\
 & w(t,\bm{X})=\int W(t,\bm{X},\bm{\xi})d\bm{\xi},\displaybreak[0]\label{eq:A.3}\\
 & \Theta(\bm{X}_{1},\cdots,\bm{X}_{N})=\prod_{i=1}^{N}\prod_{j>i}^{N}\theta(|\bm{X}_{ij}|-\sigma),\quad\bm{X}_{ij}=\bm{X}_{i}-\bm{X}_{j},\label{eq:A.4}
\end{align}\end{subequations}
\noindent and $D^{N}$ is the $N$-times direct multiple of $D$.
Substituting \eqref{eq:factorize} into \eqref{eq:density}, the density $\rho$ is expressed in terms of $w$ as
\begin{equation}
\rho(t,\bm{X})=\frac{mN}{\Phi(t)}w(t,\bm{X})Y(t,\bm{X}).\label{eq:Y}
\end{equation}
The correlation function $g_{2}$ {in \eqref{eq:g_g2}} is defined as
\begin{subequations}\label{eq:PairFunc}
\begin{align}
 & g_{2}(t,\bm{X}_{1},\bm{X}_{2})\nonumber \\
= & \frac{m^{2}N(N-1)}{\Phi(t)}\frac{w(t,\bm{X}_{1})w(t,\bm{X}_{2})}{\rho(t,\bm{X}_{1})\rho(t,\bm{X}_{2})}\nonumber \\
 & \quad\times\int_{D^{N-2}}w(t,\bm{X}_{3})\cdots w(t,\bm{X}_{N})\Theta_{(1,2)}(\bm{X}_{1},\cdots,\bm{X}_{N})d\bm{X}_{3}\cdots d\bm{X}_{N},\label{eq:g2}
\end{align}
where
\begin{equation}
\Theta_{(1,2)}(\bm{X}_{1},\cdots,\bm{X}_{N})=\prod_{i=1}^{N}\prod_{j>\max(i,2)}^{N}\theta(|\bm{X}_{ij}|-\sigma).\label{eq:Theta12}
\end{equation}
Note that
\begin{equation}
\Theta(\bm{X}_{1},\cdots,\bm{X}_{N})=\theta(|\bm{X}_{12}|-\sigma)\Theta_{(1,2)}(\bm{X}_{1},\cdots,\bm{X}_{N}),\label{eq:A.7}
\end{equation}
\end{subequations}by \eqref{eq:A.4} {and \eqref{eq:Theta12}.
By \eqref{eq:Y} with \eqref{eq:A.1},
$\rho$ can be regarded as a functional of $w$ and, if invertible, vice versa. Hence, $\Phi$ and $g_{2}$ can also be regarded as functionals of $\rho$.

Below, the argument $t$ is suppressed unless confusion is expected, and the summation convention for repeated indices is not used.

\paragraph{\textbf{Case I}}
Assume that the system under consideration is axially symmetric about the $z$-axis. The geometry of $D$ must also be axially symmetric about the $z$-axis. 
Then, $w(\bm{X})=w(\mathrm{R}\bm{X})$ holds by the axial symmetry,
where $\mathrm{R}$ is a rotation matrix about the $z$-axis.
Since $D$ is invariant under the rotation $\mathrm{R}$,
$\Theta$ is also invariant under the rotation by \eqref{eq:A.4}.
Thus, the axial symmetry of $w$ propagates to $Y$ and $\rho$,
see \eqref{eq:A.1} and \eqref{eq:Y}.

Let $(P_i,\phi_i,z_i)$ be the cylindrical coordinates of $\bm{X}_i$ 
and let $\mathrm{R}_i$ be the rotation matrix that moves the position $\bm{X}_i$ to $\bm{Y}_i$ with the cylindrical coordinates $(P_i,2\phi_1-\phi_i,z_i)$. The new position $\bm{Y}_i=\mathrm{R}_i\bm{X}_i$ is a mirror image of $\bm{X}_i$ with respect to the plane spanned by $\bm{X}_1$ and the $z$-axis. If $\bm{X}_1$ is on the $z$-axis, simply put $\phi_1=0$. Since the relative distances do not change under the transformations $\mathrm{R}_2,\cdots,\mathrm{R}_N$, $|\bm{Y}_{ij}|=|\bm{X}_{ij}|$ and
$\Theta_{(1,2)}(\bm{X}_1,\cdots,\bm{X}_N)=\Theta_{(1,2)}(\bm{X}_1,\bm{Y}_2,\cdots,\bm{Y}_N)$ hold. The integral in \eqref{eq:g2} can therefore be transformed as follows:
\begin{align}
&\int_{D^{N-2}}w(\bm{X}_{3})\cdots w(\bm{X}_{N})\Theta_{(1,2)}(\bm{X}_{1},\bm{X}_{2},\cdots,\bm{X}_{N})d\bm{X}_{3}\cdots d\bm{X}_{N}\notag \\
=&
\int_{D^{N-2}}w(\bm{X}_{3})\cdots w(\bm{X}_{N})\Theta_{(1,2)}(\bm{X}_1,\bm{Y}_2,\cdots,\bm{Y}_N)d\bm{X}_{3}\cdots d\bm{X}_{N}\notag \\
=&
\int_{D^{N-2}}w(\bm{X}_{3})\cdots w(\bm{X}_{N})\Theta_{(1,2)}(\bm{X}_1,\bm{Y}_2,\cdots,\bm{Y}_N)d\bm{Y}_{3}\cdots d\bm{Y}_{N}\notag \\
=&
\int_{D^{N-2}}w(\bm{Y}_{3})\cdots w(\bm{Y}_{N})\Theta_{(1,2)}(\bm{X}_1,\bm{Y}_2,\cdots,\bm{Y}_N)d\bm{Y}_{3}\cdots d\bm{Y}_{N}.\label{eq:g2RHS}
\end{align}
Note that the integration range does not change under the change of variables made at the third equality and that
the rotational invariance of $w$ is used at the last equality.
Using the rotational invariance of $\rho$ and $w$ again on the right-hand side of \eqref{eq:g2}, 
\begin{equation}
g_2(\bm{X}_1,\bm{X}_2)=g_2(\bm{X}_1,\mathrm{R}_2\bm{X}_2),\label{eq:g2even}
\end{equation}
is obtained.
That is, $g_2(\bm{X}_1,\bm{X}_2)$ is even with respect to $\phi_2-\phi_1$.

\paragraph{\textbf{Case II}}
Assume that the system under consideration is invariant under a translation in the $z$-direction. The geometry of $D$ must also be invariant under the same translation. 
By a similar argument to Case I, 
$w$, $Y$, and $\rho$ are invariant under a translation in the $z$-direction.

Now let $\mathrm{S}_i$ be the translation that moves the position $\bm{X}_i$ to $\bm{Z}_i$ with the cylindrical coordinates $(P_i,\phi_i,2z_1-z_i)$. The new position $\bm{Z}_i=\mathrm{S}_i\bm{X}_i$ is a mirror image of $\bm{X}_i$ with respect to the plane normal to the $z$-axis containing $\bm{X}_1$. Since the relative distances do not change under the transformations $\mathrm{S}_2,\cdots,\mathrm{S}_N$, $|\bm{Z}_{ij}|=|\bm{X}_{ij}|$ and
$\Theta_{(1,2)}(\bm{X}_1,\cdots,\bm{X}_N)=\Theta_{(1,2)}(\bm{X}_1,\bm{Z}_2,\cdots,\bm{Z}_N)$ hold. Hence, by the transformation similar to \eqref{eq:g2RHS},
\begin{equation}
g_2(\bm{X}_1,\bm{X}_2)=g_2(\bm{X}_1,\mathrm{S}_2\bm{X}_2).\label{eq:g2zeven}
\end{equation}
That is, $g_2(\bm{X}_1,\bm{X}_2)$ is even with respect to $z_2-z_1$.






\section*{Acknowledgments}
The present work has been supported in part by the JSPS KAKENHI Grant
(No.~22K03923) and the Kyoto University Foundation.
The authors thank Masanari Hattori for his comments to the first draft of this paper.









\medskip
Received xxxx 20xx; revised xxxx 20xx; early access xxxx 20xx.
\medskip


\begin{thebibliography}{99}



\bibitem{BLPT91}
\newblock N. Bellomo, M. Lachowicz, J. Polewczak and G. Toscani,
\newblock \emph{Mathematical Topics in Nonlinear Kinetic Theory II},
World Scientific, Singapore, 1991. 


\bibitem{B95}
\newblock L. Boltzmann, 
\newblock \emph{Lectures on Gas Theory}, Dover edition, New York, 1995, Part I, Chap. II, Sec. 18.

\bibitem{BGM17}[10.1103/PhysRevE.96.042117]
\newblock J. J. Brey, M. I. Garcia de Soria and P. Maynar,
\newblock \doititle{Boltzmann kinetic equation for a strongly confined gas of hard spheres},
\newblock \emph{Phys. Rev. E}, \textbf{96} (2017), 042117.
%
\bibitem{CIP94} [10.1007/978-1-4419-8524-8]
\newblock C. Cercignani, R. Illner and M. Pulvirenti,
\newblock {\emph{The Mathematical Theory of Dilute Gases}}, Springer, New York, 1994.

%
\bibitem{CC95}
\newblock S. Chapman and T. G. Cowling,
\newblock {\emph{The Mathematical Theory of Non-Uniform Gases}}, 3rd ed. Reprint, Cambridge University Press, New York, 1995, Sec.~16.4.
%
\bibitem{CGHJK96}[10.1007/BF02124750]
\newblock R. M. Corless, G. H. Gonnet, D. E. G. Hare, D. J. Jeffrey and D. E. Knuth,
\newblock \doititle{On the Lambert W function},
\newblock \emph{Advances in Computational Mathematics}, \textbf{5} (1996), 329--359.




\bibitem{DVK21}[10.1017/9781139025942]
\newblock J. R. Dorfman, H. van Beijeren and T. R. Kirkpatrick,
\newblock {\emph{Contemporary Kinetic Theory of Matter}}, Cambridge University
Press, Cambridge, 2021.

\bibitem{E72}[10.1016/B978-0-08-016714-5.50016-6]
\newblock D. Enskog, 
\newblock Kinetic theory of heat conduction, viscosity,
and self-diffusion in compressed gases and liquids, 
\newblock \emph{Kinetic Theory}, Vol. 3, S. G. Brush ed., Pergamon Press, Oxford, Part 2, 1972, 226--259.

\bibitem{F97} [10.1063/1.869247]
\newblock A. Frezzotti, 
\newblock \doititle{A particle scheme for the numerical solution of the Enskog equation}, \newblock \emph{Phys. Fluids}, \textbf{9} (1997), 1329--1335. 

\bibitem{F97b} [10.1016/S0378-4371(97)00143-X]
\newblock A. Frezzotti,
\newblock \doititle{Molecular dynamics and Enskog theory calculation of one dimensional problems in the dynamics of dense gases}, 
\newblock \emph{Physica A}, \textbf{240} (1997), 202--211.


\bibitem{F99}[10.1016/S0997-7546(99)80008-9]
\newblock A. Frezzotti, 
\newblock \doititle{Monte Carlo simulation of the heat flow in a dense hard sphere gas}, \newblock \emph{Eur. J. Mech. B/Fluids}, \textbf{18} (1999), 103--119.



\bibitem{G49}[10.1002/cpa.3160020403]
\newblock H. Grad,
\newblock \doititle{On the kinetic theory of rarefied gases}, 
\newblock \emph{Communications on Pure and Applied Mathematics}, \textbf{2} (1949), 331--407.


\bibitem{HN06}[10.1088/0951-7715/19/6/001]
\newblock S.-Y. Ha and S. E. Noh,
\newblock \doititle{New \emph{a priori} estimate for the Boltzmann--Enskog equation}, \newblock \emph{Nonlinearity}, \textbf{19} (2006), 1219--1232.

\bibitem{HTT22}[10.1063/5.0091390]
\newblock M. Hattori, S. Tanaka and S. Takata,
\newblock \doititle{Heat transfer in a dense gas between two
parallel plates}, 
\newblock \emph{AIP Advances}, \textbf{12} (2022), 055323.

\bibitem{HCB64}
\newblock J. O. Hirschfelder, C. F. Curtiss and R. B. Bird,
\newblock \emph{Molecular Theory of Gases and Liquids}, John Wiley \& Sons,
New York, 1964, Sec.~9.3. 

\bibitem{J14}[10.1090/conm/618/12351]
\newblock P. M. Jordan,
\newblock \doititle{A Note on the Lambert W-Function:
Applications in the Mathematical and Physical Sciences},
\newblock \emph{Contemporary Mathematics},
\newblock \textbf{618} (2014), 247--263. 


\bibitem{K38}
\newblock E. H. Kennard, 
\newblock \emph{Kinetic Theory of Gases}, McGraw--Hill,
New York, 1938, Sec. 27.



\bibitem{MGB18}[10.1007/s10955-018-1971-7]
\newblock P. Maynar, M. I. Garcia de Soria and J. J. Brey,
\newblock  \doititle{The Enskog equation for confined elastic hard spheres}, 
\newblock \emph{J. Stat. Phys.}, \textbf{170} (2018), 999--1018. 

\bibitem{P98}
\newblock B.~Perthame,
\newblock Introduction to the collision models in Boltzmann's theory,
\newblock \emph{Modeling of collisions}, Gauthier-Villars, Gap, 1998, 141--176.


\bibitem{R78}[10.1007/BF01011771]
\newblock P. Resibois, 
\newblock \doititle{H-theorem for the (modified) nonlinear Enskog equation}, 
\newblock \emph{J. Stat. Phys.}, \textbf{19} (1978), 593--609.

\bibitem{S07}[10.1007/978-0-8176-4573-1_3]
\newblock Y. Sone, 
\newblock \doititle{\emph{Molecular Gas Dynamics}}, Birkh\"{a}uer,
Boston, 2007. Supplement is available from \url{http://hdl.handle.net/2433/66098}.

\bibitem{T23}[10.3934/krm.2023025]
\newblock S. Takata, 
\newblock \doititle{On the thermal relaxation of a dense gas described by the modified Enskog equation in a closed system in contact with a heat bath},
\newblock \emph{Kinetic \& Related Models}, (2023).


\bibitem{TM80}
\newblock C.~Truesdell and R.~G.~ Muncaster,
\newblock \emph{Fundamentals of Maxwell's kinetic theory of a simple monatomic gas: treated as a branch of rational mechanics}, Academic Press, New York, 1980.

\bibitem{VE73} [10.1016/0031-8914(73)90372-8]
\newblock H. van Beijeren and M. H. Ernst, 
\newblock \doititle{The modified Enskog equation}, 
\newblock \emph{Physica}, \textbf{68} (1973), 437--456. 

\bibitem{WLRZ16} [10.1017/jfm.2016.173]
\newblock L. Wu, H. Liu, J. M. Reese and Y. Zhang,
\newblock \doititle{Non-equilibrium dynamics of dense gas under tight confinement}, 
\newblock \emph{J. Fluid Mech.}, \textbf{794} (2016), 252--266.

\end{thebibliography}
\end{document}